\documentclass{article}

\usepackage[dblblindworkshop, final]{neurips_2025}

\workshoptitle{MATH-AI}

\usepackage[utf8]{inputenc} % allow utf-8 input
\usepackage[T1]{fontenc}    % use 8-bit T1 fonts
\usepackage{hyperref}       % hyperlinks
\usepackage{url}            % simple URL typesetting
\usepackage{booktabs}       % professional-quality tables
\usepackage{amsfonts}       % blackboard math symbols
\usepackage{nicefrac}       % compact symbols for 1/2, etc.
\usepackage{microtype}      % microtypography
\usepackage{xcolor}         % colors

\usepackage[pdftex]{graphicx} % the pdftex option was suggested by the workshop template so the below is commented out
\usepackage{changes} 
\usepackage{comment}
\usepackage{caption}
\usepackage{tablefootnote}
\usepackage[normalem]{ulem}
\usepackage{amsmath}
\usepackage{markdown}       % adding this in case Wendy still needs it

\usepackage{mdframed}
\usepackage{placeins}
\newmdenv[
  backgroundcolor=gray!15, % light gray (adjust number for darker/lighter)
  linecolor=black,         % border color
  leftmargin=1em,
  rightmargin=1em,
  skipabove=1em,
  skipbelow=1em,
]{smallerbox}

\newenvironment{smallermdframed}
  {\begin{smallerbox}\scriptsize} % reduce font size
  {\end{smallerbox}}
  
\title{Adaptive Coopetition: Leveraging Coarse Verifier Signals for Resilient Multi-Agent LLM Reasoning}

\author{
\textbf{Rui Jerry Huang}\\
{\small Basis Independent Silicon Valley}\\
{\small \texttt{ruihuang15352019@gmail.com}}\\
\And
\textbf{Anastasia Miin}\\
{\small Pacific Collegiate School} \\
{\small \texttt{anatasiamiin9@gmail.com}}\\
\And
\textbf{Wendy Liu}\\
{\small The Harker School}\\
{\small \texttt{blossomwliu@gmail.com}}\\
\And
\textbf{Lei Ding~\thanks{Corresponding author: lding25@ucsc.edu. Lei acknowledges that intellectual property rights related to this work are held by the three first authors, in recognition of their contributions and passion for the research.}}\\
{\small University of California, Santa Cruz}\\
 {\small \texttt{lding25@ucsc.edu}}\\
}
\date{August 2025}

\begin{document}

\maketitle

\begin{abstract}
Inference-time computation is a critical yet challenging paradigm for enhancing the reasoning performance of large language models (LLMs). While existing strategies improve reasoning stability and consistency, they suffer from notable limitations: self-correction often reinforces the model's initial biases, and Multi-Agent Collaboration (MAC) often fails due to the lack of efficient coordination mechanisms, leading to collective errors. Although high-performing verifiers can detect reasoning errors, making them reliable requires substantial training. To address these challenges, we introduce a novel inference-time framework - \textbf{Adaptive Coopetition (AdCo)} - in which LLM agents utilize \textbf{an adaptive, UCB-based `coopetition' mechanism}. At each round, agents leverage coarse verifier signals to determine whether to collaborate or compete, further iteratively refine their reasoning based on peer feedback. Without relying on high-performance verifiers, our adaptive strategy achieves significant performance gains on mathematical reasoning benchmarks, yielding \textbf{a 20\% relative improvement} over baselines on the more challenging dataset. Our approach remains robust and consistent in terms of accuracy under different sample sizes and configurations. This adaptive, signal-guided `coopetition' framework enhances reasoning robustness by leveraging both
model knowledge diversity and reasoning trace measure, while also promoting uncertainty-driven exploration, especially when participants have comparable capabilities. From this perspective, our work offers a fresh lens on inference-time computation and paves the way for more resilient multi-agent LLM systems. Our code is available at \url{https://github.com/AdCo-Research/adaptive-coopetition}.
\end{abstract}

\section{Introduction}

Nowadays, LLMs exhibit strong reasoning capabilities but remain limited in certain scenarios due to inherent pre-trained knowledge scope~\cite{mirzadeh2025, yan2025}. Although model scaling and self-correction techniques further extend their capabilities, these approaches are either computationally expensive or prone to self-bias. To address these limitations, multi-agent frameworks emerge to facilitate collective intelligence among LLM agents through coordinated orchestration. A good case in point is the use of debate-based systems~\cite{du2024multiagent, liang2024} and autonomous orchestration frameworks~\cite{wu2024autogen}. However, this line of work often suffers from reasoning collapse, stemming from rigid strategies and reasoning contamination from low-quality peer feedback. To mitigate this, many methods were proposed: leveraging strong verifiers to evaluate outputs~\cite{lifshitz2023mav, wang2024math}, and optimizing multi-agent architecture and reasoning processes~\cite{zhou2024mass,lee2025evolving,tran2025,qimutual}. Unfortunately, these methods often lack inference-time adaptability and either require extensive training or assume a symmetric role for each agent, limiting their practicality for deployment at inference time.

To overcome these challenges, we propose Adaptive Coopetition - a lightweight inference-time, multi-round multi-agent framework that enhances collective reasoning through adaptive decision-making guided by coarse verifier signals. Specifically, after one step of reasoning, each agent employs a coarse verifier to evaluate the current reasoning trace from multiple perspectives, producing what we term "verifier signals". Using these signals, AdCo applies a revised Upper Confidence Bound (UCB) algorithm~\cite{auer2002} to let each agent decide whether to collaborate (absorb a peer’s reasoning trace) or compete (invite peer criticism). With the strategy determined, agents engage in peer-to-peer (P2P) interactions and asynchronously refine their reasoning based on peer feedback. This design deliberately isolates low-quality reasoning traces~\cite{zhang2024chain, qiu2024towards} and iteratively improves the reasoning before integrating it into the cluster, thereby enhancing reasoning quality and mitigating reasoning collapse~\cite{pan2025multiagent}.

Experiments on mathematical datasets, particularly the more challenging DeepMath-103K~\cite{he2025} in terms of model capacity, demonstrate the effectiveness of our approach. The best-performing heterogeneous AdCo cluster outperforms both the State-of-the-Art (SOTA) LLMs and conventional multi-agent frameworks by approximately 20\% in accuracy, while maintaining consistently strong performance across different data scales. Further ablation studies underscore the necessity of key components in AdCo, reinforcing our belief that AdCo offers a practical and effective solution that enhances collective reasoning.

\section{Adaptive coopetition} %% Methodology section

\begin{figure}[htb]
    \centering
    \includegraphics[width=0.9\textwidth]{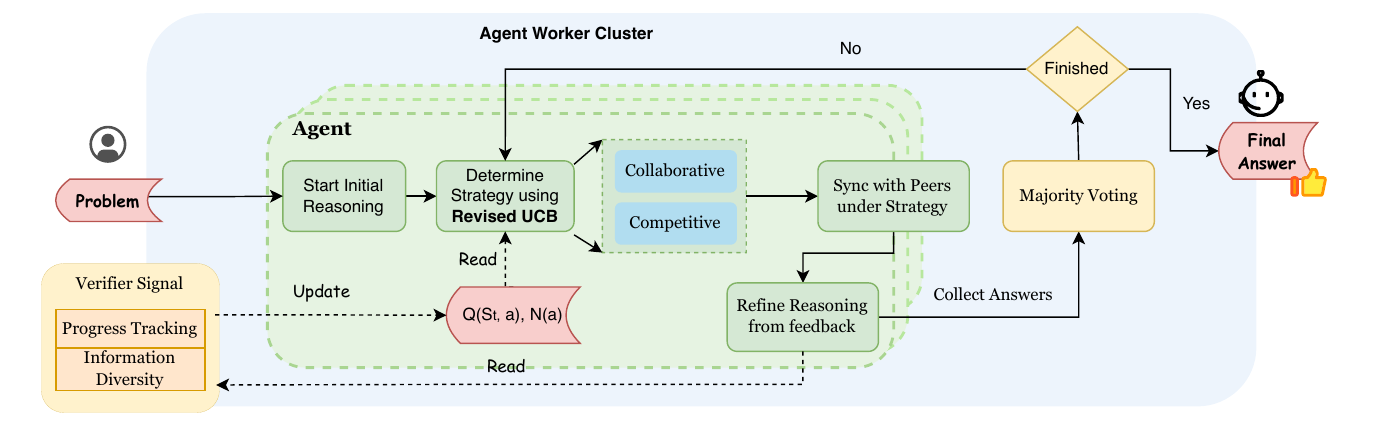}
    \caption{Overview of adaptive coopetition} 
    \label{fig:Figure 1}
\end{figure}

Figure~\ref{fig:Figure 1} illustrates how AdCo Worker Cluster solves problems through multi-round optimization. At each round, worker agents advance reasoning by one step and determine their strategy—collaboration or competition—via a UCB-based algorithm guided by verifier signals. Formally, we estimate coarse verifier signals by different reasoning trace measures: reasoning progress (via process reward~\cite{prm2025}), the diversity of reasoning trace (via semantic similarity of reasoning trace~\cite{estornell2024multi}), their weighted combination. Then, the chosen measure is used in the revised UCB algorithm to decide the strategy for the current round, prompting agents to exchange feedback with peers selectively and refine the original reasoning. This process repeats until a final solution is reached through a majority-vote algorithm~\cite{chen2025harnessingmultiplelargelanguage}. The following sections detail our core components.

\paragraph{Coarse verifier signals:} Coarse verifier signals refer to verifier outputs of moderate precision in estimating reasoning progress and reasoning trace diversity at inference time. High-precision verifiers often require substantial resources to train, and obtaining a sufficiently accurate verifier can be infeasible. Interestingly, our empirical results show that even mediocre-quality signals from coarse verifiers can still serve the intended purpose under AdCo: filter out bad or inconsistent feedback while amplifying good and consistent ones in the reasoning process.

\paragraph{Model Diversity:} Model diversity is introduced in AdCo through worker cluster configurations, aiming to reduce the risk of static debate dynamics, wherein the debate procedure directly converges to the majority opinion~\cite{estornell2024multi}. AdCo supports two model configurations: homogeneous (the same LLM model is used across all agents) and heterogeneous (different LLM models are used within the cluster). The heterogeneous configuration promotes diversity by incorporating distinct LLM models and resulting in broader pre-trained knowledge, as further evidenced by our experiments.

\paragraph{Low-quality feedback isolation:} We use a customized filter mechanism and peer-to-peer communication to prevent reasoning collapse caused by the dissemination of unqualified information~\cite{zhang2024chain, qiu2024towards}. In the collaborative strategy, agents choose the highest-scoring feedback to merge with to avoid regressing in solution quality. In the competitive strategy, agents isolate low-quality critique by requesting feedback only from the highest-scoring peer agents.

\paragraph{Iterative adaptive coopetition:} AdCo models each agent's problem-solving process as a Markov decision process. At the round \(t\), state $s_t$ is the current agent's reasoning trace. Our action space is defined as $ A \equiv \{c_0, c_1\}$, where $c_0$ is to collaborate and $c_1$ is to compete. Given the chosen action $a_t$ at round $t$, state transition is deterministic: $T(s_{t+1}|s_t, a_t) \equiv 1$. Reward $r(s_t, a_t) \in [-1,1]$ is measured by the change in the estimation value of coarse verifier signals.

Essentially, our revised UCB algorithm serves as the action policy $\pi(s_t)$, formulated as a variant of the multi-armed bandit problem~\cite{auer2002} in which rewards are assumed to be independent and identically distributed according to an unknown distribution with unknown expectation \(\mu_t\). Inspired by UCT \cite{10.1007/11871842_29}, we replace the state-independent exploitation term in UCB with a heuristic approximation that includes $s_t$.  Specifically, the chosen action $a_t$ is the candidate action $a$ that maximizes:
\begin{equation} \label{eq:1}
    UCB'(s_t,a)=Q(s_t,a)+C\times\sqrt{\frac{\ln N}{N(a)}}, a \in A
\end{equation} 
where \(Q(s_t, a)\) is the estimated payoff of action candidate $a$ at state $s_t$, \(N\) is the total number of executed actions, and \(N(a)\) is the number of times that action candidate \(a\) has been executed so far. We then measure $Q(s_t, a)$ by the average verifier signal value changes caused by action candidate \(a\):
\begin{equation}
    Q(s_t,a)=\frac{\sum_{i<t} \Delta V(s_i, a)}{N(a)}, a \in A
\end{equation}
where $\Delta{V(s_i, a)}$ is the change of verifier signal estimation at state \(s_i\) where the chosen action is $a$. More algorithm details refer to \ref{appendix:algorithm}.

%\begin{equation}
%    Q(s_t,a)=\frac{\sum_{a|s} \Delta V(s_t)}{N(a)}, a \in \{c_0, c_1\}
%\end{equation}
%where \(V(s_t)\) is the verifier signal estimation at state \(s_t\). $\Delta V (s_t)$ is measured by the mean of incremental changes in verifier signal value when action \(a\) is chosen. Formally, \(V(s_t)\) is a weighted combination of process reward and information diversity estimation.

\section{Experiments}
\label{sec:Experiments}
We evaluate AdCo's performance on GSM8K~\cite{cobbe2021training}, GSM8K-Symbolic~\cite{mirzadeh2025} and \textit{DeepMath-103K}\cite{he2025}. Preliminary tests for the chosen models reveal a clear performance saturation of AdCo on the former two, as shown in~\ref {appendix:gsms}. Consequently, we focused on the more challenging \textit{DeepMath-103K} dataset, exploring multiple data scales and assessing (1) the effectiveness of the iterative adaptive coopetition strategy; (2) the effect of low-quality feedback isolation using coarse verifier signals; and (3) the impacts of model diversity. Further details are provided in \ref{appendix:datasets}.
 
Using the Microsoft AutoGen framework~\cite{wu2024autogen}, we set up a heterogeneous Agent Worker Cluster using three LLMs: DeepSeek/DeepSeek-v3-0324~\cite{deepseekai2025}, Google/Gemma-3-27b-it~\cite{gemma2025}, and GPT-4o~\cite{openai2024}. We employ reasoning progress as the verifier signal. Qwen2.5-Math-PRM-7B~\cite{prm2025} is used as the verifier model, and its output Process Reward (PR) serves as the verifier signal value. In Equation~\ref{eq:1}, we empirically choose \(C=\sqrt{1.5}=1.22\). Algorithmic details are given in \ref{appendix:algorithm} and \ref{app:implementation_details}.

We compared AdCo against two baseline categories using the same LLMs: (1) individual LLMs with self-correction mechanisms and (2) a plain multi-agent debate approach representing multi-agent collaboration: either collaborate or compete with appropriate peers. We also evaluated AdCo in both homogeneous and heterogeneous settings to assess the impact of model diversity. See \ref{app:baselines} for details.

\subsection{Performance evaluation}

\begin{figure}[ht!]
    \centering
    \begin{minipage}[b]{0.48\textwidth}
        \centering
        \includegraphics[width=\textwidth]{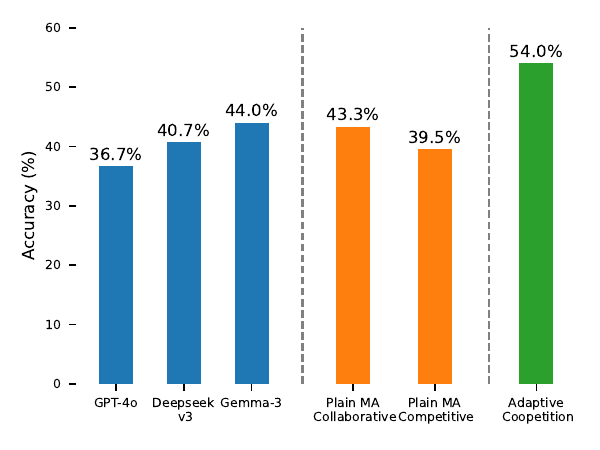}
        \caption{AdCo shows clear improvement over baselines.}
        \label{fig:Figure 2}
    \end{minipage}
    \hfill
    \begin{minipage}[b]{0.48\textwidth}
        \centering
        \includegraphics[width=\textwidth]{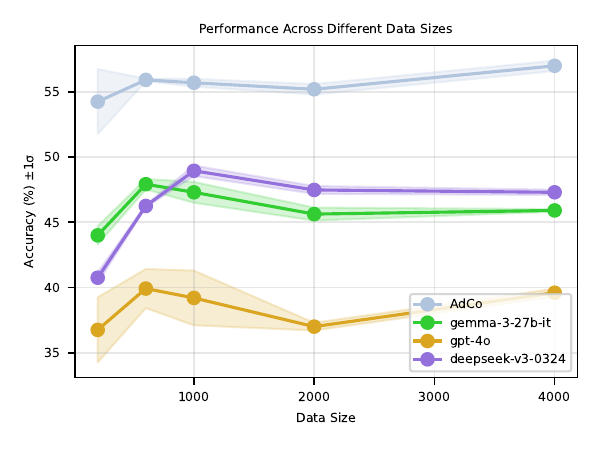}
        \caption{AdCo shows stability (STDEV < 1\%) from 600-4000 data points.}
        \label{fig:Figure 3}
    \end{minipage}
\end{figure}

\paragraph{Accuracy \& stability:} We measured accuracy using the percentage of correct final answers and stability using the standard deviation across runs. As shown in Figure~\ref{fig:Figure 2} and Figure~\ref{fig:Figure 3}, AdCo improved the accuracy from 37\%--44\% (across individual and plain multi-agent baselines) to 54\%. Moreover, the standard deviation remained low (<1\%) across various dataset sizes, indicating consistently robust performance.  

\paragraph{Model diversity:} AdCo also performs better under the heterogeneous configuration than homogeneous ones, highlighting the positive impact of model diversity. In contrast, the observed accuracies of homogeneous setups were: 52\% with $3\times$ DeepSeek-V3-0324, 51\% with $3\times$ Gemma-3-27B-IT, and 42\% with $3\times$ GPT-4o—all falling short of 54\% accuracy achieved by the heterogeneous counterpart.

\paragraph{Efficiency:} We measured efficiency using the number of successful switches from incorrect to correct answers using each strategy. In AdCo, agents are more likely to switch from incorrect to correct answers than vice versa, showing its effectiveness in guiding agents toward meaningful progress. For instance, under collaborative strategies, at 2{,}000 samples, agents made 1{,}016 switches from incorrect to correct answers, compared to only 102 switches from correct to incorrect.

\subsection{Ablation study}

\paragraph{Revised UCB-based action policy:}

Replacing the revised UCB with a simple flipping rule—where agents collaborate only when the PR exceeds 0.5 and compete otherwise—led them to make nearly three times as many corrections from incorrect to correct decisions (1,401 vs. 509 under UCB’ at 1{,}000 samples) while yielding lower accuracy (54.08\% vs. 55.70\%). These results confirm that UCB effectively leverages verifier signals to guide agents toward better decisions. See Table~\ref{table:1} for details.

\paragraph{Impact of agent capability:}

We tested AdCo with stronger models to assess whether it improves the accuracy beyond the already high accuracy of the baselines. Using $3 \times$ Qwen/QWQ-32B (standalone accuracy: 74.75\%), AdCo improved accuracy to 80.5\%, showing that even high-accuracy models benefit from AdCo. Replacing Gemma-3-27B-IT with Qwen/QWQ-32B in our current configuration yielded no significant gain (52.25\%), likely because majority voting diluted its influence. These findings suggest AdCo achieves the best relative performance improvement when agents have comparable capabilities and diverse reasoning styles. 

\section{Limitations \& future improvements}
\label{sec:limitations}
Despite promising preliminary results, we plan to introduce the following improvements in the future:

%\paragraph{Richer verifier signals} Our current implementation uses PR feedback as a proxy for the coarse verifier signal. We plan to develop more advanced quantitative measures of information diversity and integrate them with progress tracking, using a more sophisticated method to construct a comprehensive verifier signal.
\paragraph{Cost-Benefit trade-off} Despite the observed accuracy improvements, the accompanying costs—in terms of token consumption, latency, and redundant API calls—are non-trivial. Therefore, more sophisticated mechanisms should be developed to mitigate unnecessary computational overhead, particularly when only marginal performance gains are expected. We anticipate that such considerations will naturally manifest in the agent’s inter-communication strategies and in the determination of appropriate stop conditions.
\paragraph{State-aware exploration} While modifying the UCB-1 exploitation term provides a reasonable heuristic approximation, the algorithm doesn't fully capture state-dependent exploration dynamics. We will enhance state-aware exploration by incorporating the agent’s reasoning state and history. This would enable the evaluation of the benefit of exploring alternative reasoning paths under the current context. This approach is capable of more effectively balancing exploration and exploitation based on the trajectory of reasoning, potentially leading to more accurate and efficient outcomes. Last but not least, we would like to bring in further theoretical analysis for our UCB-1 variation to strengthen the case of our adapted heuristics. 
\paragraph{Weighted result aggregation} The majority voting mechanism diminished the impact of stronger agents in heterogeneous settings, as evidenced when a weaker model is replaced by a comparably stronger one. This indicates that our current aggregation strategy may under-utilize high-performing agents. We plan to explore alternatives to majority voting, such as confidence-weighted or performance-based aggregation, which may better leverage the strengths of high-performing agents.
\paragraph{Strategy-specific parameter tuning} Currently, each LLM client is configured with its default hyperparameters due to practical constraints of the AutoGen framework on the non-OpenAI models. This limitation prevents us from adapting parameters such as temperature and sampling rate to better optimize reasoning performance. In future work, we plan to conduct additional trials to enable parameter tuning across models, to improve reasoning performance and efficiency.
\paragraph{Extended experiments} As tool-augmented and structure-aware methods have demonstrated strong effectiveness in agentic applications, we are planning to incorporate tool-use agents and search/planning variants into our baselines to examine whether the performance gains follow a consistent trend. Meanwhile, considering PRM's role in the proposal, it is necessary to discuss how the performance changes when PRM accuracy is intentionally degraded or adversarially perturbed.
\paragraph{Lightweight architectures \& expansion to broader domain} Furthermore, we will explore the adoption of lightweight-trained or distilled agent models to make the framework more accessible in resource-constrained environments. We also plan to extend the framework to other reasoning-intensive domains beyond mathematics, such as scientific discovery and legal analysis, to evaluate its versatility and robustness.

\section{Conclusion and future works}
In this paper, we introduced Adaptive Coopetition, a lightweight inference-time multi-round, multi-agent framework that enhances LLM multi-step reasoning through self-evolution with peer feedback from adaptive collaboration and competition. AdCo adopts a reinforcement learning-based reflection for adaptive strategic selection, using a modified two-armed UCB-1 algorithm guided by coarse verifier signals. Experiments demonstrate that AdCo significantly outperforms self-correction standalone LLM and conventional multi-agent baselines in reasoning accuracy, stability, and strategy efficiency. Future improvements include state-aware exploration along reasoning trajectories, weighted result aggregation, strategy-specific parameter tuning, lightweight architectures for resource-limited settings, and expansion to broader domains (see ~\ref{sec:limitations}). Overall, we expect AdCo to enhance inference-time reasoning via adaptive strategy selection, while producing diverse reasoning traces with the verifier signals to inform future training and extend its impact to broader reasoning domains.

\begin{ack}
We gratefully acknowledge Professor Yang Liu for the conceptual formalization of Multi-LLM Debate, which inspired and motivated this study. We also thank Ph.D. student Yaxuan Wang for valuable assistance in refining and finalizing the diagram designs. We further appreciate the reviewers’ insightful feedback, which will inform and strengthen our future research. This work received no external funding, and the authors declare no competing interests or conflicts of interest.

%Use unnumbered first level headings for the acknowledgments. All acknowledgments
%go at the end of the paper before the list of references. Moreover, you are required to declare
%funding (financial activities supporting the submitted work) and competing interests (related financial activities outside the submitted work).
%More information about this disclosure can be found at: \url{https://neurips.cc/Conferences/2025/PaperInformation/FundingDisclosure}.

%Do {\bf not} include this section in the anonymized submission, only in the final paper. You can use the \texttt{ack} environment provided in the style file to automatically hide this section in the anonymized submission.
\end{ack}

\bibliographystyle{plain}
\bibliography{references} % Link to your .bib file

\newpage

\appendix

\section{Appendix and supplemental materials}

\subsection{Algorithm derivation}
\label{appendix:algorithm}
As shown in Equation~\ref{eq:1}, determining the next action strategy of a worker agent -- to compete or collaborate with appropriate peers -- is equivalent to maximizing its chosen reasoning trace measure. This setting resembles a traditional multi-armed bandit problem, where the Upper Confidence Bound (UCB) algorithm~\cite{auer2002} selects an arm $a$ to maximize the accumulated reward according to

\begin{equation}
  UCB(a) = \bar{X}_a + c \sqrt{\frac{\ln N}{n_a}}
\end{equation}

where $\bar{X}_a$ is the mean reward of arm $a$, $n_a$ is the number of times arm $a$ has been pulled, $N$ is the total number of pulls, and $c$ is a  exploration hyperparameter. The first exploitation term encourages exploiting actions with high observed rewards, while the second exploration term incentivizes exploring less frequently used actions.  

Here, the key distinction between the traditional UCB algorithm and our problem framing is: the reward in our case -- defined as the change in the verifier signal value after executing an action -- is state-dependent. This motivates drawing inspiration from a UCB variation applied to the tree search space (UCT)~\cite{10.1007/11871842_29}, which extends it to sequential decision processes over structured state spaces. Correspondingly, the action selection in UCT at each state $s$ is given by  

\begin{equation}
   UCT(s,a) = Q(s,a) + c \sqrt{\frac{\ln N(s)}{N(s,a)}} \label{uct}
\end{equation}

where $Q(s,a)$ denotes the estimated action-value at state $s$, $N(s)$ is the visit count of state $s$, and $N(s,a)$ is the count of selecting action $a$ from $s$. According to Equation~\ref{uct}, the stateful nature of UCT is evident: both the exploitation and exploration terms depend on the current state $s$.

To adapt UCB for AdCo, we revised the original algorithm by replacing its state-independent exploitation term with the state-dependent term $Q(s_t,a)$ in Equation \ref{eq:1}, and leaving the state-dependent exploration term as future work (See \ref{sec:limitations}). 

To measure $Q(s_t,a)$, our hypotheses are as follows:
\begin{itemize}
    \item The estimated payoff of action candidate $a$ at state $s_t$ is proportional to the average of measurable reasoning progress and information diversity increases, which reflects on $Q(s_t,a)$:
\begin{equation}
    Q(s_t,a) \propto \frac{\sum_{i<t} \Delta \text{Progress}(s_i, a)}{N(a)}, Q(s_t,a) \propto \frac{\sum_{i<t} \Delta \text{Diversity}(s_i, a)}{N(a)}
\end{equation}
    \item The estimated payoff of action candidate $a$ at state $s_t$ grows proportionally with the weighted combination of reasoning progress and the degree of information diversity gains: \begin{equation}
    Q(s_t,a) \propto \frac{\sum_{i<t} \Delta \text{Progress}(s_i, a) \odot \Delta \text{Diversity}(s_i, a)}{N(a)}
\end{equation}
where $\Delta \text{Progress}(s_i, a)$ measures the reasoning progress at state $s_i$ when the chosen action is $a$, and $\Delta \text{Diversity}(s_i, a)$ captures the resulting increase in information diversity when the chosen action is $a$, $\odot$ is the weighted combination operator.
\end{itemize}

Since we focus on reasoning progress and assume that PRM offers a rough estimate of reasoning progress, the revised UCB can be simplified as follows:
\begin{equation} %\label{eq:1}
    UCB'(s_t,a)=\frac{\sum_{i<t} \Delta PR(s_i, a)}{N(a)}+C\times\sqrt{\frac{\ln N}{N(a)}}, a \in \{c_0, c_1\}
\end{equation}

\subsection{GSM8K and GSM8K-Symbolic}
\label{appendix:gsms}

%In addition to \textit{DeepMath-103K}, we also evaluated \textit{GSM8K} and \textit{GSM8K-Symbolic} dataset. 
\textit{GSM8K}~\cite{cobbe2021training} is a dataset of 8.5K high-quality, grade-school--level math problems. It features high linguistic diversity while relying on relatively simple mathematical concepts. Each problem requires between 2 and 8 steps to solve, typically involving a sequence of elementary calculations with basic arithmetic operations ($+$, $-$, $\times$, $\div$). The dataset is carefully curated, with fewer than 2\% of problems containing critical errors, and each problem is designed to be relatively unique, ensuring both quality and diversity.

\textit{GSM8K-Symbolic}~\cite{mirzadeh2025} is characterized by its templated problem structure based on the GSM8k dataset, and enables the systematic generation of different problems from a single template by varying numerical values. This mitigates the risk of pattern matching or memorization, which can inflate performance metrics on benchmarks with a limited number of fixed examples. Consequently, this dataset can provide a more reliable measure of an LLM's mathematical reasoning capabilities, compared to the original GSM8k dataset.

To evaluate whether the \textit{GSM8K} and \textit{GSM8K-Symbolic} datasets are suitable for our experiment, we assessed the performance of the following models on these datasets: DeepSeek/DeepSeek-v3-0324, Google/Gemma-3-27b-it, and GPT-4o, as well as AdCo in a heterogeneous setup using these three models. The results on GSM8K-Symbolic (with similar results observed for GSM8K) are summarized in below Table~\ref{table:2}:

\begin{table*}[ht!]
\centering
\begin{tabular}{|p{0.2\linewidth} c c c |}
  \hline
  & 200 & 1000 & 5000 \\
  \hline
  gemma-3-27b-it & $86.25\% \pm 0.4\%$  & $85.25\% \pm 0.5\%$ & $86.35\% \pm 0.1\%$ \\
  \hline
  gpt-4o & $91.5\% \pm 0.5\%$ & $92.50\% \pm 0.3\%$ & $91.94\% \pm 0.4\%$\\
  \hline
  deepseek-v3-0324 & $89.00\% \pm 2.1\%$ & $91.10\% \pm 0.7\%$ & $91.26\% \pm 0.1\%$ \\
  \hline
  AdCo & $89.75\% \pm 1.1\%$ & $92.58\% \pm 0.1\%$ & $91.84\% \pm 0.3\%$ \\
  \hline
\end{tabular}
\caption{Performance evaluation on GSM8K Symbolic dataset}
\label{table:2}
\end{table*}
As shown in the above table, each base model already achieves $\sim 90\%$ accuracy for the GSM8K Symbolic dataset. This suggests that the underlying patterns of the GSM8K series are largely captured by the chosen models. Therefore, they leave little room to push the capability boundary of the underlying LLM with these datasets, which drives us to choose a more challenging dataset without such performance saturation. 

\subsection{DeepMath-103K}
\label{appendix:datasets}

\textit{DeepMath-103K} \cite{he2025} is a large-scale mathematical reasoning dataset released in April 2025, due to its distinctive characteristics:

\begin{itemize}
\item \textit{Unique Data Acquisition}: Unlike many open-source math datasets that predominantly repackage well-known, pre-formatted problems from standardized sources such as AIME\cite{patel2024aime} and AMC\cite{hendrycks2measuring}, \textit{DeepMath-103K} curates problems from more diverse and less-structured origins. For example, it extracts and reformulates problems from community-driven platforms like Math StackExchange into a clean, well-structured question--answer format. This results in a broader and more original problem distribution, significantly reducing overlap with prior datasets and encouraging generalizable reasoning.

\item \textit{Verifiable Answers}: Each problem includes a final, rule-verifiable answer that facilitates automated correctness checks, making the dataset well-suited for evaluating the accuracy and stability of our AdCo across multiple baselines.

\item \textit{Rigorous Decontamination}: The dataset underwent a comprehensive decontamination process to remove any overlap with established math benchmarks such as \textit{MATH}, \textit{Minerva}, \textit{AIME}, and \textit{OlympiadBench}, making it a trustworthy resource for evaluating true generalization.
\end{itemize}

Preliminary tests of the chosen models achieved only 36.7\%–44.0\% accuracy, highlighting their limited pre-trained knowledge and the substantial performance gap that AdCo can address. To ensure unbiased evaluations, we randomly sampled a scaled-size 200, 400, 600, 1{,}000, 2{,}000, and 4{,}000 problems with numeric answers from the \textit{DeepMath-103K}. All samples were selected through uniform random sampling without replacement to avoid selection bias.

\subsection{Baseline configurations} \label{app:baselines}

We evaluate Heterogeneous AdCo against two categories of baselines, as well as its Homogeneous counterpart:
\begin{itemize}
    \item \textit{Individual LLMs with Self-Correction}: Each model operates independently with iterative self-refinement (DeepSeek-v3, Gemma-3, GPT-4o).
    \item \textit{Plain Multi-Agent (AutoGen)}:
    \begin{itemize}
        \item Collaborative-only setting: All agents collaborate based on peers' partial solutions to refine their reasoning without AdCo.
        \item Competitive-only setting: All agents critique peers' partial solutions to refine their reasoning without AdCo.
    \end{itemize}
    \item \textit{Homogeneous AdCo}: 3 identical LLM agents (e.g., $3\times$ DeepSeek) applying AdCo under the same model type used in the corresponding heterogeneous setting.
\end{itemize}

\subsection{Implementation details} \label{app:implementation_details}

\subsubsection{Verifier model}
Qwen2.5-Math-PRM-7B \cite{prm2025} is chosen as our verifier model, because 1) it can evaluate intermediate reasoning steps and not just the final answer 2) it shows suitable performance identifying errors in standard benchmarks (such as ProcessBench, etc.) and Best-of-N evaluations. Moreover, we evaluated its performance on several different datasets, and found that the reported PR accuracy on the DeepMath dataset is relatively low (\textless 50\%), making it a good candidate to act as a coarse signal provider. 

\subsubsection{LLM client setting} 
\label{sssec:LLM configuration}
In the experiment, each LLM client was configured using the default AutoGen hyperparameter settings. While tuning these parameters for each LLM client would be preferable—and we initially attempted to do so—we ultimately kept defaults for consistency. For example, under the competitive strategy, we considered increasing the temperature to encourage exploration of alternative reasoning paths and raising the sampling rate to identify better and pursue high-confidence candidates. Conversely, under the collaborative strategy, lower temperatures and reduced sampling would be more appropriate.

However, we were unable to implement further hyperparameter tuning due to the practical constraints of our chosen framework. Confidence scores are only supported by OpenAI or some self-hosted models, excluding the other models in our experiments, and the AutoGen framework does not allow configurable sampling rates without source code modification — forcing sequential exploration that is prohibitively slow and costly at scale.

\subsubsection{Worker agent design}
\begin{figure}[htb]
    \centering
    \includegraphics[width=\textwidth]{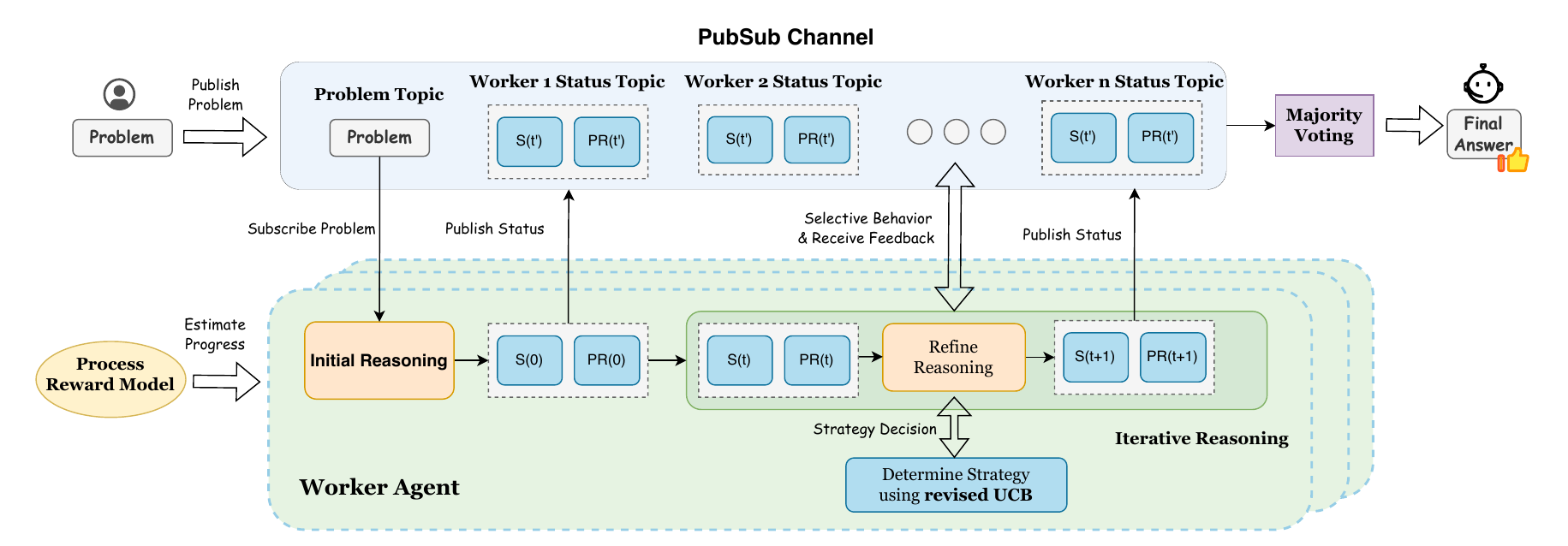}
    \caption{Worker agent architecture: relying on Pub/sub channel to exchange information within the worker cluster, each agent individually carries out initial reasoning and continues iterative reasoning until the cluster consensus is reached via majority voting.} %Each agent in the Agent Worker Cluster begins with initial reasoning, selects a strategy using Upper Confidence Bound (UCB), synchronizes with peers, and iteratively refines its reasoning based on feedback. The system aggregates responses through majority voting and updates verifier signals based on progress and diversity, encouraging continual improvement across the Agent Cluster as agents refine reasoning.}
    \label{fig:Figure 4}
\end{figure}

To support efficient self-evolution and filter out low-quality reasoning in the cluster, a general asynchronous message-driven architecture with selective peer-to-peer communication has been built on top of the AutoGen framework for AdCo, shown in Figure~\ref{fig:Figure 4}. The following discussion concentrates on how a typical worker agent — say, Agent A — iteratively self-evolves.

\paragraph{Problem reception and initial reasoning}

Initially, Agent A subscribes to the problem topic on the shared Pub/Sub channel and receives the published problem once available. It then invokes the corresponding LLM to generate its first reasoning step. Next, Agent A queries the Process Reward Model, obtaining $PR(0)$ (verifier signal) for the current partial solution $S(0)$. Both $S(0)$ and $PR(0)$ are then published to Agent A's work status topic on the shared channel. Eventually, the initial reasoning state $S(0), PR(0)$ is persisted as a cluster-accessible topic, serving as the starting point for its subsequent reasoning. Unlike future rounds $t > 0$, the initial reasoning step only generates the initial PR(0) without involving interactions with other peers in the cluster.

\paragraph{Iterative reasoning}

After the initial step, Agent~A enters a cycle of iterative reasoning: stepping forward from $S(t), PR(t)$ to $S(t+1), PR(t+1)$ until answer convergence. At each round $t$, Agent~A decides its action strategy using the revised UCB algorithm, which takes the performance gain $\Delta PR = PR(t-1) - PR(t-2), t > 1$ at the previous round as input, and then executes the chosen action accordingly:

\begin{itemize}
    \item Competitive: Agent~A selects the peer agent with the highest average performance (excluding itself) to critique the current partial solution. The average performance is defined as the cumulative PRs up to round $t$ normalized by the number of rounds, i.e., $\frac{\sum_{i=0}^{t-1} PR(i)}{t}$. Then, the selected peer interacts directly with Agent~A via AutoGen's peer-to-peer communication channel: it retrieves Agent~A’s partial solution $S(t-1)$, critiques it using the prompt \ref{fig:Prompt 3}, and sends the feedback back directly to Agent~A. Agent~A then integrates this critique feedback with the prompt \ref{fig:Prompt 4} to refine its reasoning at round $t$.

    \item Collaborative: Agent~A retrieves all $S(t-1), PR(t-1)$ from other peers via the shared work status topic, but only incorporates the $S(t-1)$ with the highest $PR(t-1)$ from peer agents into the prompt \ref{fig:Prompt 2} at the current round $t$.
\end{itemize}

After this round, the updated partial solution $S(t)$ and its corresponding $PR(t)$ are published back to their own worker status topic.

\paragraph{Convergence check}
At the end of round $t$, a monitoring daemon reviews the worker status topic to access outputs from all worker agents and determine whether convergence has been reached. If so, outputs are aggregated through majority voting to produce the final answer (see \ref{sssec:majority voting}). 

\paragraph{Iterating until convergence} If convergence has not been reached, a new round $t+1$ begins, with the agent’s state updated to $S(t), PR(t)$, following the aforementioned logic. This cycle continues iteratively until all agents converge on a final answer.

\subsubsection{Majority voting and final answer determination}
\label{sssec:majority voting}
%\paragraph{Majority voting and final answer determination:} \mbox{}\\
AdCo uses the following criteria to determine whether a final answer has been reached:

\begin{itemize}
    \item All agents have reached a final answer after at least two rounds; \textbf{or}
    \item A quorum of agents have converged on the same final answer, and more than 5 rounds have been completed. (We chose 5 rounds to ensure adequate debate among the three agents while also keeping costs manageable. As future work, we plan to conduct further testing to identify the optimal number of debate rounds.); \textbf{or}
    \item If the number of rounds exceeds 20, the final answer is determined via majority voting among the agents. We limited the rounds to 20 to manage inference time costs.
\end{itemize}

\subsection{Worker agent prompts} 
This section includes the prompts each worker agent uses to 1) perform initial reasoning; 2) refine reasoning using peer feedback under collaborative strategy; 3) critique a peer’s partial response under competitive strategy; 4) refine reasoning via peer critique under competitive strategy. 

\begin{figure}[htbp]
\begin{smallermdframed} 

You are assisting with a math reasoning problem by providing the next step in the solution process. Your explanation should be clear, concise, and generate only one extra step.\par

\text{\#Steps:}\par

\begin{enumerate}
    \item Analyze the given math problem and the previous steps provided.
    \item Create a clear summary of the previous steps and include them in your response.
    \item Identify the next logical step to progress the solution.
    \item Explain the step clearly, showing how it advances the problem-solving process.
    \item If this step leads to the final answer, present it using the format: \texttt{The answer is \#\#\#\# [numerical answer].}
\end{enumerate}

\text{\#Output Guidelines:}\par
\begin{itemize}
    \item Create a clear summary of the previous steps, and include only one additional step in the response.
    \item Use the final answer format if the solution is complete: \texttt{The answer is \#\#\#\#[numerical answer]}
    \item Keep your response under 100 words.
\end{itemize}

\text{\#Notes:}\par
\begin{itemize}
    \item Focus on clarity and logical reasoning.
    \item Ensure continuity by building directly from previous steps.
\end{itemize}

Now given the following math problem and previous steps, add the next step.\par  

\begin{verbatim}
Problem: {content}\n
Previous steps: {prev_steps}\n
\end{verbatim}

\end{smallermdframed}
\caption{Initial reasoning prompt}
\label{fig:Prompt 1}
\end{figure}

\begin{figure}[htbp]
\begin{smallermdframed} 
You are a math reasoning assistant. Your role is to solve a problem step by step by integrating the best parts of two given partial solutions.

\text{\#Steps:}
\begin{enumerate}
    \item Carefully read and understand the math problem.
    \item Review both partial solutions thoroughly.
    \item Extract and combine the strongest reasoning from each partial solution to create a unified solution.
    \item If the final answer hasn’t been reached, provide only the next logical step.
\end{enumerate}

\text{\#Output Format:}
\begin{itemize}
    \item Rewrite the combined solution. If the final answer is still incomplete, provide just one additional step per response.
    \item Keep your response under 100 words.
    \item If this step solves the problem, present the answer as: \texttt{The answer is \#\#\#\#[numerical answer]}
\end{itemize}

Now given the following math problem, two partial solutions, please generate the next step.  

\begin{verbatim}
Problem: {content}\n
solution_1: {solution_1}\n 
solution_2: {solution_2}\n 
\end{verbatim}

\end{smallermdframed}
\caption{Collaborative strategy - refine reasoning using peer feedback}
\label{fig:Prompt 2}
\end{figure}

\begin{figure}[htbp]
\begin{smallermdframed} 
Your task is to review a partial solution to a math problem and identify any errors.

\text{\#Steps:}
\begin{enumerate}
    \item **Understand the Problem**: read and comprehend the math reasoning problem.
    \item **Review the Partial Solution**: Check for mistakes in logic or calculation.
    \item **Critique**: explain any errors found clearly.
\end{enumerate}

\text{\#Output Format:}
\begin{itemize}
    \item Provide a concise critique to the partial solution; do not provide the final answer in the response.
    \item Keep your response under 100 words.
\end{itemize}

\text{\#Notes:}
\begin{itemize}
    \item Focus on accuracy in identifying mistakes.
    \item Ensure your explanation is clear and to the point. 
\end{itemize}

Now given the following math problem and partial solution, please carefully inspect the solution and point out any mistakes.  

\begin{verbatim}
Problem: {content}\n
Partial solution: {peer_response}\n
\end{verbatim}

\end{smallermdframed}
\caption{Competitive strategy - provide critique on peer’s partial response}
\label{fig:Prompt 3}
\end{figure}

\begin{figure}[htbp]
\begin{smallermdframed} 
Your task is to review a partial solution and its critique for a math reasoning problem, correct any errors, and provide the next correct step in the solution.

\text{\#Steps:}
\begin{enumerate}
    \item **Understand the problem**: read and interpret the math problem. 
    \item **Review the partial solution**: identify any mistakes or gaps.
    \item **Evaluate the Critique**: assess the critique’s accuracy. 
    \item **Address the Critique**: replace the partial solution with a corrected solution. If the final answer hasn’t been reached, provide only the next logical step.
\end{enumerate}

\text{\#Output Format:}
\begin{itemize}
    \item Add only one step per response.
    \item Clearly explain your reasoning.
    \item If reaching the final answer, use the format: \texttt{The answer is \#\#\#\#[numerical answer]}
    \item Keep your response under 100 words.
\end{itemize}

Now given the following math problem, previous steps and critique, please carefully consider the critique and correct any mistakes as the next step. 

\begin{verbatim}
Problem: {content}\n
Previous steps: {prev_steps}\n
Critique: {critique}\n
\end{verbatim}
\end{smallermdframed}
\caption{Competitive strategy - refine reasoning using peer critique}
\label{fig:Prompt 4}
\end{figure}

\FloatBarrier

\subsection{Ablation study - the revised UCB vs. simple flipping rule}
\begin{table*}[ht!]
\centering
\begin{tabular}{|p{0.2\linewidth} c c c |}
  \hline
  & 200 & 600 & 1000 \\
  \hline
  UCB & $54.25\% \pm 2.5\%$ & $55.92\% \pm 0.1\%$ & $55.70\% \pm 0.3\%$ \\
  \hline
  Flipping & $52.00\% \pm 4.2\%$ & $54.58\% \pm 0.1\%$ & $54.08\% \pm 0.3\%$ \\
  \hline
\end{tabular}
\caption{Performance comparison of the revised UCB vs. simple flipping}
\label{table:1}
\end{table*}

%%%%% For the time being we will not include the paper checklist since past workshops mostly did not require it

\newpage

%%% The checklist below may not be needed for workshop papers. The input line below can be commented out to remove the checklist.
%\input{neurips submission checklist}

\end{document}